\documentstyle{article}
\begin{document}
\begin{center}
{
The harmony, reflection and other principles of complex systems
}\\
{D.B. Saakian}\\
{Yerevan Physics Institute,\\
Alikhanian Brothers St. 2, Yerevan 375036, Armenia.}
\end{center}

\begin{abstract}

A set of general physical principles is proposed as the structural basis for
the theory of complex systems. First the concept of harmony is analyzed and its 
different aspects are uncovered. Then the concept of reflection is defined and illustrated 
by suggestive examples. Later we propose the principle of (random) projection of
symmetrically expanded prereality as the main description method of complex systems.

\end{abstract}

\vspace{5mm}

During the last decade 
complex phenomena attracted a serious attention [1-5].
In [8-9] we suggested to consider  complex phenomena on the basis of Random 
Energy Model (REM)[10]. Here we try to explore the idea of reflection, 
introduced in [8], as well as to introduce the concept of harmony.\\
It is our intuitive idea  that complex system arises due and during the harmonic multilevel 
reflection, when subjective reality transforms into objective reality.\\
Let us clarify these notions.
\section{The Harmony}
The concept of harmony is one of the fundamental concepts of philosophy since the
ancient Greece. In situations when experimental verification of theory is impossible
physicists often appeal to the internal beauty of theory. We believe intuitively, that 
system consisting of many parts or carrying different features is beautiful if these parts
are in harmony. Let us clarify and specify the concept of harmony, as the main aspect of 
beauty for (physical) systems.\\
1.Symmetry\\
This is the first concept of harmony. Circle and square are harmonic objects. Here the 
number of  symmetric operation is comparable with the order of the group.
Within the common viewpoint  a theory is interesting, when all parts of Lagrangian have the same symmetry 
group. Especially  important is the case of local gauge invariance (in principle, other 
cases with local symmetric property are also possible, as for models of Self-Organized 
Criticality). Physicists consider symmetric theories since the 
discovery of the hydrodynamics and classical electrodynamics. \\
It appears  that local symmetric property corresponds to information processing for the
 case of spin-glasses and neuron networks. Indeed, let us consider a Hamiltonian for $N$ 
 discrete spins $s_i=\pm 1$,
\begin{equation}
\label{e1}
H=-\sum_{i,j}j_{ik}s_{i}s_{ik}
\end{equation} 
This Hamiltonian is symmetric under discrete transformations $\eta_i=\pm 1$
\begin{eqnarray}
\label{e2}
s_i\to s_i\eta_i\nonumber\\
j_{ik}\to j_{ik}\eta_{i}\eta_{k}
\end{eqnarray} 
If one constructed couplings $j_{ik}$ to have  $\{s_i\}=\{s^1_i\}$ as a vacuum
 (minimum  energy) configuration for this Hamiltonian, then for any other configuration
$\{s^2_i\}$ one can easily formulate Hamiltonian with this new configuration as a 
vacuum:
\begin{eqnarray}
\label{e3}
\eta_i=s^1_is^2_i\nonumber\\
j^2_{ik}\to j_{ik}\eta_{i}\eta_{k}
\end{eqnarray}
So local gauge invariance assumes some internal mechanisms for information processing,
 information transmission from the couplings $j$ to ground state configuration of spins 
 $s_i$.\\
What about classical fields? Perhaps here one can consider information transmission from
 state at $t=0$ to field state at other moments of time.\\
So the obvious  harmonic property-local symmetric property means some information 
processing mechanisms, which is less obvious. \\ 
2.Variation principle.\\
G. Galileo believed that planetary orbits are circular. We know, that in mechanics the 
harmony is deeper than the principle of minimal action. The variation
principle constraint means, that the motion is perfect in some
sense. It is interesting the situation, when the same solution could be found from the 
different variation principles or at same constraints. \\
3.a.Parametric resonance\\
Let us consider hierachial systems. There are controller parameters (couplings), and 
simple dynamic variables. Let as assume, that both level of hierarchy can be described by
 some feature (collective parameter), characterizing the total state of hierarchy level. 
 When those two parameters coincide, system is becoming perfect in some sense. A simple 
 example is paramagnetic
resonance with frequency as collective parameter. When we change length with frequency
 equal to   resonance frequency of oscillator happens a resonance. Another (less famous) 
 example is Nishimori line in spin-glasses [11]. In Hamiltonian (1) there are spins 
 $s_i=\pm 1$, as well as couplings $j_{ik}$ 
Couplings are ($\pm 1$) or real with asymmetric distribution. This distribution is 
$\rho(j)/\rho(-j)\sim e^{-\beta_nj}$, where $\beta_n$ looks like as inverse
 temperature. Let us consider statistical mechanics of (1) at inverse temperature $\beta$. 
 When two temperatures coincide, the model  becomes solvable and has some peculiar 
 properties.
This phenomena later has been explored to other symmetry groups also. \\
I mentioned two examples, where collective parameter is real number
(temperature, frequency). There should exist situation, where
controller parameter is $\pm 1$- ("warm" and "cold") or even word (too complicated 
complex systems).
\\We mentioned the case of hierarchic system. 
The situation with several reflections of the same system  is similar.
If it is possible to define the objectives for those levels   
those levels
(hierarchic or reflection)
in certain way, then paramagnetic resonance corresponds to the situation,
when those objectives coincide. In other words-appearance and the objective should correspond
 to each other.\\
4.Multi (logical) modeling of the same system.\\
In mathematics it is known, that the same theory sometimes may correspond to different 
logical models. In physics sometimes we can exactly formulate theory in different ways 
(velocity or vorticity equations in hydrodynamics) or solve it in different ways.
In quantum mechanics of Hydrogen atom it is possible to separate variables in different 
ways. This is the reason of peculiar (perfect) properties. The similar is the situation with 
2d exact solvable models (different ways of Bethe anzats).\\
Let us remember two equivalent formulations of Random Energy Model REM.\\
a. One has $2^N$ energy levels $E_i$ with the independent distribution 
$\rho(E)\sim \exp(-\frac{E^2}{N})$. 
One defines partition at fixed values of energies 
$Z(\{E_i\})=\sum_i\exp(-\beta E_i)$ and observable free energy 
$F=<\ln Z(\beta,\{E_i\})>_{E_i}$.\\
b.Let us consider N spins $s_i$ interacting via the hamiltonian 
$$H=-\sum_{1\le i_1<i_2..i_p\le N}j_{i_1..i_p}s_{i_1}..s_{i_p}$$ here p is 
some large number. In this (mean field) hamiltonian there are
 $M=\frac{N!}{p!(N-p)!}$ terms. Couplings have normal distribution with proper norm
 $<j>=0, <j^2>2M=N$.\\
Derrida proved, that at thermodynamical limit $(N\to \infty)$ these models are  equivalent.\\
5.Edge of chaos\\
Let us consider the situation, when there are two phases for system. Instead of choosing one 
of them system chooses the "golden middle", the border between them. The fact is that 
this point is highly symmetric, it was well known for second order phase transition between
 ferromagnetic and paramagnetic phases. More intriguing is the case of border between 
 chaotic (undeterministic, anomalous) and ordered (deterministic) phases. First it was 
 found, that in dynamical systems at the border between chaotic and ordered regimes 
 system is maximally adaptable, so could evolve in  an optimal way. Perhaps the life was 
 born here [7,8].\\
 In the case of Random Energy Model(REM) this border between ferromagnetic (when
  $<j>\ne 0$) and spin-glass phases has another interpretation-the threshold of errorless 
  coding (the border of true and false!). Again information transmission takes place from
   the couplings to the vacuum configuration. In this system there are paramagnetic phase 
   ( no  spontaneous magnetization), ferromagnetic phase (spontaneous global magnetization), 
  and spin glass phase ( there is some magnetization, which changes randomly from    site
   to site). \\
 Really paramagnetic and ferromagnetic phases are both deterministic and so are closer 
 to each other, than to spin-glass phase. The paramagnetic-spin glass border is also 
 interesting. It was known more than a decade that it is possible to solve strings according
  to ideas of A.M. Polyakov at dimensions $d\le 1$. If one analytically continues 
  formulas for $d>1$, then at $d\ge1$ string
 transforms from paramagnetic phase to SG.
 So the famous $d=1$ barrier for bosonic string corresponds to
 this case [13]. If some theory
could be handled at some continuous parameter, then it is very
interesting the point at the border of anomaly.\\
6.Modalities.\\
 R.S.Ingarden, A. Kossakowski and M. Ohya introduced the concept of modality as 
"possible non- categorical attitudes to reality". From their point of view space, potential 
energy, information  and
entropy are stairs of the hierarchical staircase. When we are constructing  statistical
mechanics of a system and calculating partition at given values of volume, temperature (entropy), number of copies (SG models with finite replica number, number of generations
in particle physics), we really are connecting this modalities into a mixture.
 In the REM modeling of 2d models this 
correspondence becomes explicit. The space volume's ($\frac{L^d}{a^d}$) logarithm 
corresponds to some entropy in REM like picture, Laplacian's correlator (potential) 
corresponds to intensity of noise. Technically this correspondence was successful due to
 logarithmic form of 2d correlator.\\ 
Is it possible
any harmony between different "stairs"? Some facts are known: the case of two copies is special, 
in esthetics there is notion of golden section. This point is very important and needs in a serious investigation, especially the harmonic ration between emergent properties during the
reflection and nonemergent properties.
 We considered 6 aspects of harmony. To go forward we should define the notion of reflection.

\section{The reflection}
Reflection is a mapping phenomena from original space to space mapped space which \\
a.Touches all the parts of system.\\
b.Conserves certain crucial  interrelations  of original system.\\
c.Compresses phase space.\\
From the point of view of the observer the original  phase space is connected with objective
 reality, as the reflected one-where as subjective reality reflected phase space.
For every mapping it is possible to introduce entropy,  possible to introduce also free 
energy in mapped space, then again we have energy on the new level. Such reflection is
 defined  as thermodynamically complete.\\
 In usual statistical mechanics configuration space is mapped into the free energy and some 
 other thermodynamic variables like magnetization. Here the crucial property, conserved 
 during the mapping, is the property to do mechanical work on macroscopic level (for the
  Q-value spin model the number of configurations is  $D_1=Q^N$, then free energy 
  $F_1\sim N\sim \ln D_1$). Both free energy and magnetization together form  the 
  subjective (thermodynamically  complete)reality. The number of degrees depends on 
  the number of couplings can vary from a few till $N\gg 1$. \\
Let us consider the variety of thermodynamical variables as configuration space for the new
mapping process: $D_2\sim N$.  If  there is a usual free energy $F_2$, then in this
 double hierarchic case $$F_2\sim \ln D_2\sim \ln N$$ So the existence of such 
 logarithmic free energies is connected with the existence of hierarchic structure, even when
  it is hidden. We assume, that this feature is crucial to have real complexity.\\ 
The second example of thermodynamic reflection is connected with multiscaling. \\
{\bf Multiscaling}.
One has some random field $p_x$ in the box with infrared and ultraviolet cutoffs a and L. 
The number of degrees in d dimensional  space is $\sim (\frac{L}{a})^d$. Let us define
$$P(q,\frac{L}{a})=\sum_x p(x)^q$$
 and the averaged one via some probability distribution 
$$\tau(q)=<\ln P(q,\frac{L}{a})>$$ 
defines the distribution of scaling dimensions. If one has scaling 
$$p_x\sim (\frac{a}{L})^{\alpha}$$
then there is a distribution of indices $\alpha$:
$$d \alpha'\rho(\alpha')(\frac{a}{L})^{-f(\alpha')}$$
Here the function f is defined
$$\alpha=\frac{d \tau(q)}{d q},f(\alpha)=\alpha q-\tau(q)$$
and $-\frac{1}{q}\ln \frac{L}{a}$ plays the role of free energy. It describes how one can 
manipulate with measuring process (averaging via different degrees of random variable).\\
{\bf Scale invariant theories}.
 Here there are no dimensional parameters in theory. If there
are infrared cutoff a and ultraviolet L, the number of degrees $D\sim (\frac{L}{a})^d$. 
After renormalization one has for free energy
$$F\sim \ln \frac{L}{a}\sim \ln D$$
or sometimes for the 2d conformal theories even $F=const$. In the case of the  Liouville model
 the existence of double reflection is almost explicit. One defines as partiion 
$$Z=\int D_g\phi e^{\frac
{1}{8\pi}\int d^2w\sqrt{\hat g}{\phi \Delta \phi +QR\phi-te^{\alpha\phi})}}$$
where $\phi(w)$ is  field in 2d space with coordinates $w$, $t,\alpha ,Q$ are 
parameters, $R$ is curvature, $\hat g $ 2d metrics, $\Delta $ is Laplacian.
We see, that partition Z is defined on configuration space of other (low level) partition
$\int d^2w\sqrt{\hat g}e^{\alpha\phi}$. \\
{\bf Random Energy Model (REM)}.
One has $D=2^N$ energy configurations, distributed with probability
$\frac{1}{N\pi}\exp[-\frac{E^2}{N}]$.
With such normalization of energy one has phase transition at finite temperature. Then for 
physical free energy $F\sim \ln D$.\\
{\bf Grand Unification}.
The scale for weak and strong interactions order of $1-100$ GEV are created by quantum 
fluctuations in scales $10^{15}$ GEV according to Grand Unification theory.\\
Free energy is a macroscopic manageable amount of motion (energy). Another generalization
of our logarithmic free energy picture corresponds to the case, when the macroscopically 
manageable result
is logarithmic from our original efforts. Such is the situation with chaotic motion.
Here the accuracy of data after long period of time is logarithmic from original accuracy.\\
In the  case of complex adaptive systems Gellman introduces the notion of schemata-highly 
compressed crucial information, which again resembles our highly compressed free energy.\\
 We see, that this kind of situation (logarithmic free energy) is not seldom in physics.\\
One can define different classes of complexity.  The most interesting case corresponds to 
situation, when subjective reality with his emergent properties exists, but becomes too dynamic. 
To describe such a situation one has  (for the case of multiscaling) to introduce even new 
free energy, so we have a new (thermodynamic) complete reflection. When subjective reality 
becomes complete it resembles objective one (for a second reflection). So we deduce nontrivial 
notion, that {\bf in this complexity phase there is a transition between subjective reality and  objective one}.
In information theory this   corresponds to the threshold of errorless decoding. Here the 
correctness of our definition is explicit-there is a meaning, 
if errorless decoding is possible. \\
The edge of chaos corresponds to this series also. This is the phase of complex adaptive 
systems (CAS). Here system has some emergent property M, originated via some external
 managing parameter J. Their connection $\frac{d M}{d J}$ is maximal at this point. 
 This property could be rigorously proved on the border spin glass phase-ferromagnetic 
 phase in REM. Here we have multiscaling. In principle other criteria of completeness also
  are possible (without  free energy) with weaker than logarithmic compression. \\
We defined harmony for hierarchial structures (controlling parameters. 
Sometimes system has several reflections (due to selforganization?). 
 If it is possible to define the essence for 
those reflections
in some way, then harmony corresponds to the situation,
when those essences coincide. In other words-appearance and essence should correspond
 to each other.
One can add this property to 3.a of harmony as a new feature:\\
3.b.Harmonic reflection\\
We have some free external parameters on highest level, no free parameters for other (lower) reflections. \\
This property allows efficiently the cutting of low levels of hierarchy and simplifying the
 life. Let us consider these aspects in physics.\\
 Symmetry aspects are most old and popular, especially in particle
 physics. But it the lowest level of symmetry. As Shreodinger mentioned, such symmetries
  concern to the symmetries (harmony) in biological systems, as a house painter's work to 
  the Rafael's pictures. Local invariance concept is deeper, but not too much. As we mentioned, 
  there is an internal property of information processing, when any combination of spins is 
  allowed. In the case of more developed information systems we already have words and nontrivial grammar.\\
Multi-modeling aspects are perhaps important for exactly solvable models, strings and 
their generalizations. Actually, a theory  theory becomes really complex only when having this property-
{\bf any deep true should has several faces}. This is somewhere connected with the 
porism notion in philosophy. In this context one can remember multiple interpretation of religious texts.\\
I would like to underline the concept of edge of chaos. Why  is it so important? In a system one
 has some controlling order parameters  e.g. $<j>\ne 0$- in the case of REM, genotype in the
  biology. As a result there arise some emergent (highly collective) properties of the system. As a rule
   system prefers to interact with the environment just via these emergent properties
    (magnetization in the case of REM, phenotype in biology). At the edge of chaos the 
	connection  between the controlling variable and emergent property ($\frac{d M}{d J}$) is maximal, 
	that's why system becomes adaptable.\\

\section{Projection of canonically (unitarly or replically ) expanded prereality}

Let us remember the replica trick in (mean field) quenched  disordered systems. We have
$\frac{d< Z^n>}{d n}=<\ln Z>$. 
We first define our theory in the normal space (an integer $n$, the annealed model),
then enlarge it to the case of any real n.\\
In this language one can introduce naturally some probabilistic picture in replica space
in accordance with the  Parisi's theory, assuming the replica symmetry breaking.\\
Such situations happen very often in physics and it is connected with  almost all revolutions 
in physics, including the relativity theory.
\\While one has been considering together different forms of (matter) motions , theory 
became contradictory. To overcome difficulties
 one explores the original phase space to some larger phase space. In this larger space everything is 
 normal, motion is given by canonical transformation in classical
  mechanics or unitary transformation in quantum theory. Then we return back to our physical
   reality, projecting this enlarged prereality and so creating
 some probabilities.\\
Well known examples are  nonrelativistic quantum mechanics with its infinite dimensional
 Hilbert space instead of the six  dimensional space of the classical mechanics,  relativistic 
 quantum theory, which can include in one state of quantized field different number of 
 particles, or string theory, where
 Polyakov began from the coordinates of strings in our space-time and then added  field of 
 2d quantum gravity.\\
Perhaps we can baptize this situation as "Principle of (random ?) projecting of canonically
 explored prereality" (the randomness appears in all cases besides the case with string's) \\
In reality we observe random processes, but while looking at system on deeper level, we 
can found more symmetric picture. This resemble the restoration of broken symmetries at 
high energies
in particle physics.\\
In some sense we created a reflection, where prereality is an objective reality, reality is
subjective one. In the case of quantum mechanics there is a highly compressed reflection, 
for the relativistic quantum theory and for the replica trick the compression is not so drastical. 
On this way randomness naturally appears in the complex phenomena.\\
We found two examples only: quantum expansion of prereality and replica-space case. It will
be interesting to find their applications to practical situations. Technically 
resemblance with quantum physics appears in some models of population dynamics or markets [13].
We hope, that connection is more than technical.

\section{Paradigms, language}
We assume, that there should be universality classes of complex phenomena. They can be 
specified by levels and type of reflections as well as the mechanism of information transmission
(language). In the case of interesting physical theories (strings,
turbulence,SG) there are local symmetries which corresponds to trivial grammar. We assume, 
that{\bf all the physical theories with logarithmic free energies and local symmetries could be modeled
(sometimes exactly) by REM}. The REM version of a theory is its compressed, its essence. 
It was second time, when one can missed a space in formulation of theory [14]. 
Really space is one of modalities, namely non-cathegorical aspects of reality[15]. 
For the REM class of complexity one can miss one modality-space. May be for another situations
it would be possible to miss another modalities. \\
For the case of language itself we have another class of universality. Here the n-length block entropy scales in sublinear way[16], which is weaker compression, than logarithm in REM case. 
It is very interesting to look for the statistical models, where free energy scales in a sublinear way $F\sim D^{\alpha}, 0<\alpha<1$.\\
Perhaps there are some harmonic situations here:  two replicas are special case for spin-glasses, it is famous also "golden section".

\section{Harmony and reflection outside of physics}
Meaning of ethics is a minimal number of reflections. Modesty means absence of reflections 
instead of reality. Real meaning of love in accordance of harmony -to have a family, babies.
 If one is having affairs  instead of having a family, he tries to replace it by its reflection 
 (only sex). \\         
Very interesting is reflection in language.
If we have some text, we can write some annotation, then the
annotation of rhe annotation. How many times? Another example is in linguistics
and it corresponds to the appearance of the  slang .
 After using of the word "not" we get  a reflection with the change of the  direction.
 Examples of (negative) reflections are situations, when a friend becomes an enemy or black-white 
photo appears in negative. A nice example of reflection is a joke, giving the simplified version of reality.
How dully could  life be, if people interacted with each other via their principles, ideals, instead of 
actions, which in fact are reflections  of those principles.  \\ 
Edge of chaos property is very important outside  physics. It is very interesting, when the same 
 system carries in different, almost opposite properties, which it reveals at different situations. 
 This is crucial feature of harmony in human relations and in esthetics. Like pure colors have 
 been crucial for the pictures of impressionists, we like it very much, when \\
a. person has explicit features, different in different aspects.  How nice is the boy (girl),  good mannered with other 
 persons and very warm with his bride (fiancee).\\
 We can observe some other features of harmony
here:\\
b.correspondence between words, actions and thinking\\
c.Perform life  with big strokes  (instead of their weak reflections)\\
d.finishing of any cycle.\\
The property b. corresponds to harmonic reflection, c. also has connection with it. Property 
d. allows to define a meaning of action, only in this case it is possible observe logical multi-
reading. \\
When system (person) supports edge of chaos principle (carries almost opposite features) together 
with {\bf purity}, it has 
a lot of energy and while  losing this property
it becomes dissipative. This property is crucial for humans, without it any serious harmony
 is impossible . It will be interseting to define it on quantitative level for other complex 
 systems. Here it it could be connected with conservation or circulation of some free energy among the hierarchy levels.
\\
In economy an example of reflection is money, which reflects some property of goods. Then 
stock reflects some properties of money. The harmony principle of reflections is crucial in 
economy-  too much stocks (absence of harmony) is the main reason for economical crush.\\          
Very interesting is our concept of transformation of subjective-objective realities in politics. It resembles 
 revolution or revolutionary situation according to Lenin. The correctness of definition 
here is much more explicit, than in physical systems.\\
  Another interesting example of harmony is the neck neck of the swan. If we compare it with the giraffe's neck we see, that the swan's one
  explicitly reveals the equality of the different fields in their war-weight and muscles, so 
  the neck soars in the air. 
 Due to curved geometrical form of neck swan is able to bite with $\exp[\alpha L]$ 
 different configurations of neck, so it carries a property 
of strong reflection.   The same is the situation with the human waist: efficient tube for the
 muscles, with the minimal amount of matter at the same time.   \\

\section{Resume}

We suggested several concepts for complex systems. The harmony and reflection are crucial 
for drastical simplification of the picture of the  universe. An observer has a small 
complexity, he can accept small amount of information. There are a lot of hierarchy levels, 
but not only we can survive now, but also people in primordial societies just due to this 
property: every hierarchy level
 carries essence of the previous (deeper one), so every one can understood the truth after it 
 is simplified on a proper way (proper times of reflection).\\ 
The number of hierarchy levels or reflections of complex system which are harmonic is the main 
characteristics of complex system.\\
In principle there could be some other principles. For example, the idea of Jaynes [1], that there is only one probability in
 reality,( so it is not clever thing to fracture it) was crucial for solving spin-glasses.\\
Another concept corresponds to emergence. As a rule different complex systems have internal 
principles (Schemata according to Gellman), but they interact with each other during their 
actions, which are consequence of those principles. Another example genotype-phenotype
pair in biology. Here genotype corresponds to external couplings in spin-glasses, cells-to spins, 
phenotype - to magnetization. Magnetizations are emergent properties (essentially collective).
Why external interaction is preferable via emergent features? An explanation could be done 
on the basis of information theory. Without any correlation the defect of magnetization could
be suppressed only via N, and in the case of emergence exponentially.\\
To model complex systems one first should clarify the number
(how many time subjective reality transforms into objective one)
and type (compression of phase space) of reflections, the degree of harmony and the 
language. Then one constructs reflection on the inverse 
(decompression of the phase space) direction according to principle of "Projection of symmetrically expanded prereality".
Main problem  is  to find different universality classes of complex phenomena as well as  harmony among 
different modalities, emergent and non-emergent properties. 
This work was supported by ISTC fund Grant A-102.

\end{document}